# Tunable interaction-induced localization of surface electrons in antidot nanostructured $Bi_2Te_3$ thin films

Hong-Chao Liu[†,#], Hai-Zhou Lu[‡,#], Hong-Tao He[†,§], Baikui Li[†], Shi-Guang Liu[†], Qing Lin He[†], Gan Wang[†,§], Iam Keong Sou[†], Shun-Qing Shen[‡,*] and Jiannong Wang[†,*]

[†]*Department of Physics, The Hong Kong University of Science and Technology,*

*Clear Water Bay, Hong Kong, China*

[‡]*Department of Physics, The University of Hong Kong,*

*Pokfulam Road, Hong Kong, China*

[§]*Department of Physics, South University of Science and Technology of China,*

*Shenzhen, Guangdong 518055, China*

[#] *These authors contributed equally to this work*

* Address correspondence to: sshen@hku.hk , phjwang@ust.hk

# ABSTRACT

Recently, a logarithmic decrease of conductivity has been observed in topological insulators at low temperatures, implying a tendency of localization of surface electrons. Here, we report quantum transport experiments on the topological insulator $Bi_2Te_3$ thin films with arrayed antidot nanostructures. With increasing density of the antidots, a systematic decrease is observed in the slope of the logarithmic temperature-dependent conductivity curves, indicating the electron-electron interaction can be tuned by the antidots. Meanwhile, the weak anti-localization effect revealed in magnetoconductivity exhibits an enhanced dominance of electron-electron interaction among decoherence mechanisms. The observation can be understood from an antidot-induced reduction of the effective dielectric constant, which controls the interactions between the surface electrons. Our results clarify the indispensable role of the electron-electron interaction in the localization of surface electrons and indicate the localization of surface electrons in an interacting topological insulator.

Topological insulators are gapped band insulators, but have gapless modes on their boundaries.[1–4] Recently, there is growing interest in many-body interactions in topological states of matter.[5–13] Despite the tremendous interest in physics driven by strong interaction, in the prototype materials of topological insulator, such as $Bi_2Se_3$ and $Bi_2Te_3$,[14–16] the interaction is thought to be fairly weak due to the large background lattice dielectric constant.[10,13] Nevertheless, an interaction-induced transport phenomenon resulting from the interplay of interaction and disorder may occur even for weak interaction as long as disorder scattering is sufficiently strong.[17–19] In two-dimensional (2D) electron gas, the effect manifests itself as a decrease in conductivity with logarithmic decreasing temperature. In particular, in sufficiently strong magnetic fields, the effect could be identified when other effects such as weak localization (WL) or weak anti-localization (WAL), which may also exhibit logarithmic temperature dependence,[19] are suppressed. The signature of the interaction effect has been observed in recent transport experiments on $Bi_2Se_3$ and $Bi_2Te_3$,[20–25] but convincing quantitative comparison is still lacking.[26] A direct way to verify the effect is to modulate the interaction between the surface electrons, then to measure the responses of the finite-temperature conductivity and magnetoconductivity. However, the modulation of interactions between electrons in a solid is difficult thus was rarely addressed.

In this article, we report quantum transport experiments by introducing antidot nanostructures in topological insulators. In finite-temperature conductivity measurements, as the antidot density changes, we observe a continuous and repeatable change in the slope of the conductivity as a function of logarithmic temperature, indicating that the denser antidots suppress the tendency of localization of the surface electrons. Meanwhile, the WAL effect manifested in the magnetoconductivity measurements, also shows a tunable localization tendency with the antidot density. It is well accepted that the Dirac fermions cannot be localized by impurities and disorders meanwhile the ordinary electrons can be easily localized in two dimensions. Our experimental data could not be understood within the framework of the existing theories. In order to explain our experimental observations the electron-electron interaction must inevitably be taken into account. We propose a modulation mechanism that antidot array induces a change in the effective dielectric constant of the system, which in turn modulates the electron-electron interaction. The experimental data are fitted reasonably well while this modulation mechanism is included in the transport theory for interacting and disordered Dirac fermions.[27] We therefore show the indispensable role of the interaction in localization of the surface electrons in disordered topological insulators.

**RESULTS AND DISCUSSION**

The thickness of $Bi_2Te_3$ thin film used in the experiment is 20 nm. A series of antidots arrayed in a periodic triangular lattice are fabricated between the voltage measuring probes of the samples, as shown in Figure 1a. For five different nanostructured samples in our experiment, the edge-to-edge distances $d$ of two neighboring antidots are 40, 90, 130, 190, and 250 nm, respectively. A smaller value of $d$ represents a larger density of antidots. The diameter of each antidot is fixed at 200 nm for all samples. Figure 1b and its inset show the scanning electron microscope (SEM) and atomic force microscopy (AFM) images, respectively, of the antidot nanostructured $Bi_2Te_3$ film with $d = 40$ nm (sample d40).

Low-temperature dependence of conductivity can reveal the tendency of localization or delocalization at low temperatures. Figure 2a shows the logarithmic temperature dependent conductivity $\sigma(T)$ curves of different antidot nanostructured samples and a no-antidot thin film sample as a reference, in the absence of magnetic field. All curves are normalized by their maximum conductivities and corresponding temperatures. Below a threshold temperature (typically ~ 10 K), it is evident that for all samples $\sigma(T)$ decreases linearly with decreasing $\ln T$, manifesting the localization tendency. This means that all electrons in both bulk and surface states tend to be localized and the samples become insulating with decreasing temperature. In Figure 2a, the linear slopes of $\sigma(T)$ curves decrease monotonically as antidots separation $d$ decreases, indicating the antidot array creates a continuous and systematic change to the localization tendency. A magnetic field applied perpendicular to the sample surface is found to strengthen the localization tendency. Taking sample d190 as an example, Figure 2b shows a series of $\sigma(T)$ curves at different magnetic fields ranging from 0 T to 2 T. With increasing magnetic field, the conductivity shifts downwards and the slopes of the $\sigma(T)$ curves become steeper. The slope can be defined qualitatively as $\kappa = (\pi h/e^2)(\partial\sigma/\partial\ln T)$: A steeper slope or a larger $\kappa$ means an enhanced localization tendency. By linear fitting the curves in Figure 2b, the obtained $\kappa$ are plotted in Figure 2d as a function of magnetic field $B$. A sharp increase of the slope at low magnetic fields ($B < 0.5$ T) and almost constant slope when $B > 0.5$ T are observed which indicates that the localization tendency of electrons is strengthened by the applied magnetic field. All samples show a similar dependence of the slope on applied magnetic field, which will be analyzed in details later in Figure 4b. Figure 2c shows $\sigma(T)$ curves of all samples with $B = 2$ T. As it can be seen, the dependence of the slop on the antidot array density is the same as the case with $B = 0$ T, i.e. the linear slopes of $\sigma(T)$ curves decrease monotonically as antidots density increases.

In order to understand the sharp increase of the slope in low-field regions in Figures 2b and 2d, the magnetoconductivity has been measured for all samples and is shown in Figure 3a. As it can be seen, magnetoconductivity, which is defined as $\delta\sigma(B) = \sigma(B) - \sigma(B = 0)$, is negative and exhibits a cusp around $B = 0$ T for all samples at 2.1 K, which is a typical signature of the WAL effect as a result of the $\pi$ Berry phase for the surface electrons in topological insulators.[28-37] It is known that the WAL effect enhances the conductivity with decreasing temperature, and follows a ln$T$ dependence in two dimensions,[19] so it gives a negative contribution to the slope observed in $\sigma(T)$ curves. However, the WAL effect can be quenched by applying a small magnetic field as is shown by the negative magnetoconductivity in Figure 3a. As a result, the quenching of WAL effect leads to the sharp increase of the slope of $\sigma(T)$ curves as well as the down shift of $\sigma(T)$ curves in low-field region, as shown in Figures 2b and 2d. On the other hand, the negative magnetoconductivity is suppressed with decreasing antidots separation $d$ as shown in Figure 3a. Using the magnetoconductivity formula for interacting Dirac fermions,[27,38] the measured magnetoconductivity curves in Figure 3a can be quantitatively analyzed. The phase coherence length $\ell_\phi$ can be obtained by the fittings and is shown in Figure 3b as a function of $d$. We note that the phase coherence length is shortened with decreasing $d$, indicating that denser antidot array tends to enhance the inelastic scattering which breaks the phase coherence and gives rise to the suppression of the magnetoconductivity in Figure 3a.

Figure 3c shows the magnetoconductivity of sample d190 measured at temperatures from 2.1 K to 12 K. By fitting the magnetoconductivity curve at each temperature, the $\ell_\phi$ dependence on $T$ can be obtained. The relationship between $\ell_\phi$ and $T$ for all the samples are plotted in Figure 3d. Empirically, $\ell_\phi$ increases with decreasing temperature according to $\ell_\phi \propto T^{-p/2}$, where the exponent $p$ is positive.[39] Fitting the data shown in Figure 3d, the obtained exponent $p/2$ for all samples are given in Table 1. In a 2D disordered metal, if the dominant decoherence mechanism is the electron-electron interaction $p = 1$ or the electron-phonon interaction $p = 3$.[19] In our case, the exponent $p$ is close to 1 for all samples and approaches to 1 as the antidot density increases, which provides a strong and explicit evidence that the decoherence due to electron-electron interaction is dominant and enhanced by antidot array.

To summarize our experimental measurements: (i) The $\sigma(T)$ decreases linearly with the logarithmic temperature, indicating the localization tendency of surface electrons. (ii) The slope $\kappa$ of $\sigma(T)$ vs. ln$T$ curves decreases when the separation of the antidots decreases, i.e. the density of antidots increases. The denser the antidot array, the weaker the localization tendency of surface electrons. (iii) The magnetoconductivity reveals the WAL effect from surface electrons. The fitting exponent $p$ in the temperature dependent phase coherent length is close to 1, indicating that the electron-electron interaction is the dominant decoherence mechanism.

These observations cannot be simply understood within the framework of the localization theory. It is known that conventional electrons will be localized at low temperatures in a 2D disordered metal.[19] On the contrary, the surface massless Dirac electrons of a topological insulator is expected to be immune to localization,[40,41] which is evidenced by the WAL-type negative magnetoconductivity in Figure 3. The electron-electron interaction becomes an indispensable ingredient to resolve the puzzle on the coexistence of localization tendency in $\sigma(T)$ and WAL in $\sigma(B)$ in the same sample. Similar localization tendency was also observed in the previous studies of $Bi_2Se_3$ and $Bi_2Te_3$ thin films,[20–25] and was suggested to be due to the interaction, but convincing quantitative comparison is still lacking.[26] So far only theory for conventional electrons was exploited for the surface electrons in topological insulators.[17–19]

In the theory for the gapless Dirac fermions of topological surface states,[27] two mechanisms contributing to the $\ln T$ dependence of the conductivity are considered, one is the quantum interference, and the other is the interplay of the electron-electron interaction and disorder scattering. For gapless surface fermions of topological insulators, the quantum interference effect gives rise to the WAL effect,[42,43] which exhibits the negative magnetoconductivity as shown in Figure 3a, and produces a negative contribution to the slope $\kappa$ of $\sigma(T)$ curves. A small external magnetic field quickly quenches the quantum interference effect, leading to a rapid increase of the slope $\kappa$ in Figures 2b and 2d. After the quantum interference is quenched when $B > 0.5$ T, the saturated slope $\kappa$ only contains the contribution from the electron-electron interaction named as $\kappa^{ee}$. For gapless Dirac fermions of topological insulators,[27]

$$\kappa^{ee} = 1 - \frac{1}{\pi}\frac{\arctan\sqrt{1/x^2 - 1}}{\sqrt{1-x^2}}, x \equiv \frac{8\pi\varepsilon_0\hbar}{e^2}v\varepsilon_r, \quad (1)$$

where $\hbar$ is the reduced Planck constant, $e$ is the electron charge, $v$ is the effective velocity of the surface fermions, $\varepsilon_0$ and $\varepsilon_r$ are the vacuum and relative dielectric constants, respectively. Equation (1) shows that $\kappa^{ee}$ could be changed by modifying either $v$ or $\varepsilon_r$. In comparison with $v$ the relative dielectric constant $\varepsilon_r$ is more likely to be changed by inducing antidot array, because it takes into account the effects of the lattice ions and valence electrons. For fixed $v$, we have calculated $\kappa^{ee}$ as a function of $\varepsilon_r$ for the surface fermions as plotted in Figure 4a. One can see that $\kappa^{ee}$ decreases monotonically with decreasing $\varepsilon_r$. A qualitative comparison of Figures 4a and 2c, where the linear slopes of $\sigma(T)$ at $B = 2$ T decrease monotonically as antidots density increases, implies that one of the effects of antidot array is to reduce the effective dielectric constant of the system and to further modulate the contribution from the electron-electron interaction to the slope $\kappa$ of $\sigma(T)$. The denser the antidot array, the smaller the relative dielectric constant.

Furthermore, in order to extract $\varepsilon_r$ from the experimental data, the measured dependence of slopes $\kappa$ of ln$T$-dependent conductivity on applied magnetic fields ($0 < B < 2$ T) are plotted in Figure 4b as symbols for all samples. These dependences are fitted using the quantum transport theory of interacting and disordered surface fermions (see solid lines in Figure 4b).[27,38] The $\varepsilon_r$ as a fitting parameter are obtained for all samples and are shown in Table 1. It is clear that $\varepsilon_r$ monotonically decreases when the density of antidot array is increased. A notable deviation occurred between the fitting curves and the measured slopes for the samples d90 and d40 may result from the size effect as the sample becomes a combination of 2D and quasi-1D conducting channels as their phase coherence lengths at 2.1 K are comparable with their antidot distances $d$, as shown in Table 1. This limits the applicability of the theory for 2D systems.

We can further compare the fitted $\varepsilon_r$ from the transport data with that evaluated from the classical Bruggeman effective medium theory expressed as[46]

$$f_a \frac{\varepsilon_a - \varepsilon_r^{em}}{\varepsilon_a + 2\varepsilon_r^{em}} + f_b \frac{\varepsilon_b - \varepsilon_r^{em}}{\varepsilon_b + 2\varepsilon_r^{em}} = 0, \qquad (2)$$

where $f_a$ and $f_b$ are the volume fraction occupied by the medium $a$ with relative dielectric constant $\varepsilon_a$ and medium $b$ with $\varepsilon_b$, respectively, and $\varepsilon_r^{em}$ is the effective relative dielectric constant of the heterogeneous mixture of medium $a$ and $b$. We set $\varepsilon_a = 1$ for antidot vacuum and extract $\varepsilon_b = 22.0$ as the intrinsic dielectric constant of our $Bi_2Te_3$ thin film by fitting curve for the no-antidot sample in Figure 4b. The $\varepsilon_r^{em}$ values of different samples calculated from Equation (2) are shown in Table 1. For comparison, the relative dielectric constants obtained by the two methods are plotted as a function of $d$ in Figure 4c. It is clear that both $\varepsilon_r$ and $\varepsilon_r^{em}$ show the same tendency when $d$ is varied although their magnitudes are different. This difference is not surprising as the classical Bruggeman effective medium theory is just a rough estimate.

The applicability of modulating the effective dielectric constant by antidot array can be evidenced by the average distance between surface electrons in quantum transport. Considering the surface states as an ideal Dirac cone, the average distance $2\sqrt{\pi}/k_F$ is about several nm for the Fermi wave vector $k_F \sim 0.1$ Å$^{-1}$. However, the quantum conductivity, especially its temperature variation at low temperatures, is contributed by electrons near the Fermi surface, not all electrons in the Fermi sea. The electron density near the Fermi surface can be roughly estimated as the product of the density of states at the Fermi energy and $k_BT$ ($k_B$ is Boltzmann constant and $T$ is temperature). Therefore, the average distance of surface electrons is estimated to be about 40 to 135 nm from 10 K to 1 K.[38] This is comparable with the size and separation of the antidots and it is reasonable to conclude that the antidot array can change the effective dielectric constant in the film for those electrons contributing to the quantum transport.

**CONCLUSIONS**

The main observation in this work is that the denser antidots in topological insulator thin films suppress the localization tendency of conduction electrons. The fitting exponent *p* in the temperature dependence of phase coherent length indicates the dominant role played by the electron-electron interaction. The phenomena can be viewed as the antidot-array-induced reduction of the effective dielectric constant of the system. These reveal that the widely-observed localization tendency in low-temperature conductivity of topological insulators results from the interplay of many-body interaction and disorder scattering. Although the surface electrons are usually considered to be insensitive to scattering by impurities or disorders, or not to be localized by disorder,[40,41] our experiments suggest that this should be reexamined when many-body interaction comes into stage.

**METHODS**

The 20 nm-thick $Bi_2Te_3$ thin film used in the experiment was grown by molecular beam epitaxy on a (111) semi-insulating GaAs substrate with an undoped 85 nm-thick ZnSe buffer layer. Standard Hall measurement reveals that the $Bi_2Te_3$ thin film has an electron carrier concentration and mobility of $1.5 \times 10^{19}$ cm$^{-3}$ and 310 cm$^2$V$^{-1}$s$^{-1}$, respectively, at 2.1 K. The $Bi_2Te_3$ Hall bar with dimension 60 $\mu$m × 15 $\mu$m was firstly fabricated using standard photolithographic processes. Then a series of antidots arranged in a periodic triangular lattice are fabricated between the voltage measuring probes with electron beam lithography and dry etching techniques. Ohmic contacts of electrodes are formed by evaporation of Cr(5 nm)/Au(100 nm). All transport measurements were conducted in a Quantum Design physical property measurement system with a 14-Tesla superconducting magnet and a base temperature of 2 K. All samples are measured in a pulse-delta mode using Keithley 6221 as the current source and Keithley 2182A as the voltage meter.

*Conflict of Interest:* The authors declare no competing financial interest.

*Acknowledgment:* This work was supported in part by the Research Grants Council of the Hong Kong under Grant Nos. 605011, 604910, SEG CUHK06 and 17304414, and in part by the National Natural Science Foundation of China under Grant No. 11204183. The electron-beam lithography facility is supported by the Raith-HKUST Nanotechnology Laboratory at MCPF (SEG HKUST08).

Table 1. The obtained parameters for all samples.[38] Different from the previous works[44,45], the spin-orbit scattering time (length) is no longer a fitting parameter in our theory for Dirac fermions[27], and $\ell$ is calculated from the semiclassical conductivity. $\ell_\phi$ is fitted from magneto-conductivity curves at 2.1 K. $p$ is fitted from relation $\ell_\phi \propto T^{-p/2}$ as shown in Figures 3c and 3d. $\varepsilon_r$ is fitted from the slope curves in Figure 4b, and $\varepsilon_r^{em}$ is calculated from the Bruggeman effective medium theory Equation (2). d190 means the edge-to-edge distance between two antidots is 190 nm.

|  | $\ell$ (nm) | $\ell_\phi$ (nm) | $p/2$ | $\varepsilon_r$ | $\varepsilon_r^{em}$ |
|---|---|---|---|---|---|
| no antidot | 20.0 | 230.2 ± 3.1 | 0.67 ± 0.04 | 22.0 ± 1.2 | 22.0 |
| d250 | 9.3 | 119.2 ± 2.5 | 0.63 ± 0.03 | 8.8 ± 0.5 | 16.5 |
| d190 | 7.4 | 109.4 ± 3.1 | 0.60 ± 0.04 | 6.4 ± 0.1 | 14.1 |
| d130 | 5.2 | 88.2 ± 2.7 | 0.57 ± 0.05 | 3.2 ± 0.2 | 10.6 |
| d90 | 2.9 | 70.1 ± 1.9 | 0.53 ± 0.05 | 1.7 ± 0.5 | 7.8 |
| d40 | 1.6 | 61.7 ± 2.3 | 0.49 ± 0.06 | 1.0 ± 0.5 | 3.2 |

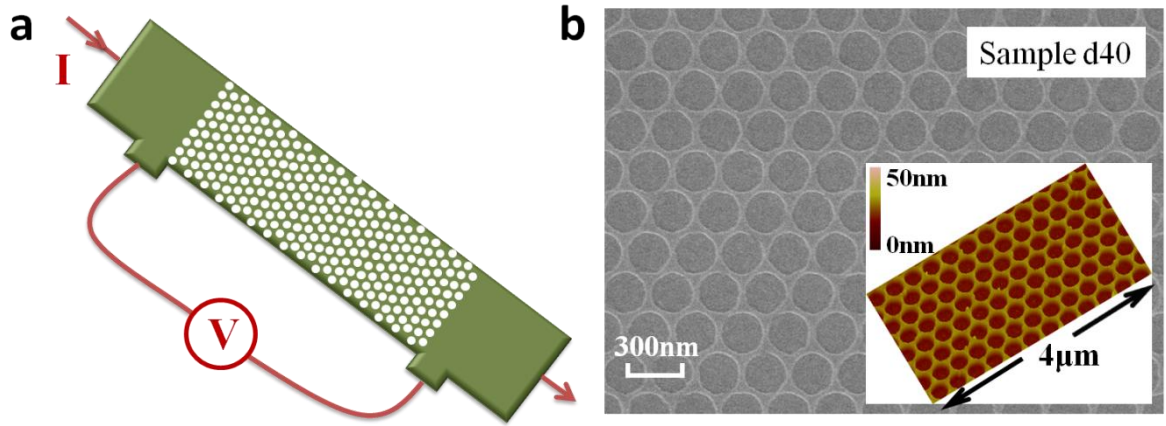

Figure 1. (a) Schematic illustration of a Hall bar sample with antidot array. (b) Scanning electron microscopy image of a $Bi_2Te_3$ thin film with antidot array, where the edge-to-edge distance between antidots $d$ = 40 nm. The inset is its atomic force microscope image, indicating the antidots depths (~25 nm) are larger than the $Bi_2Te_3$ film thickness (20 nm).

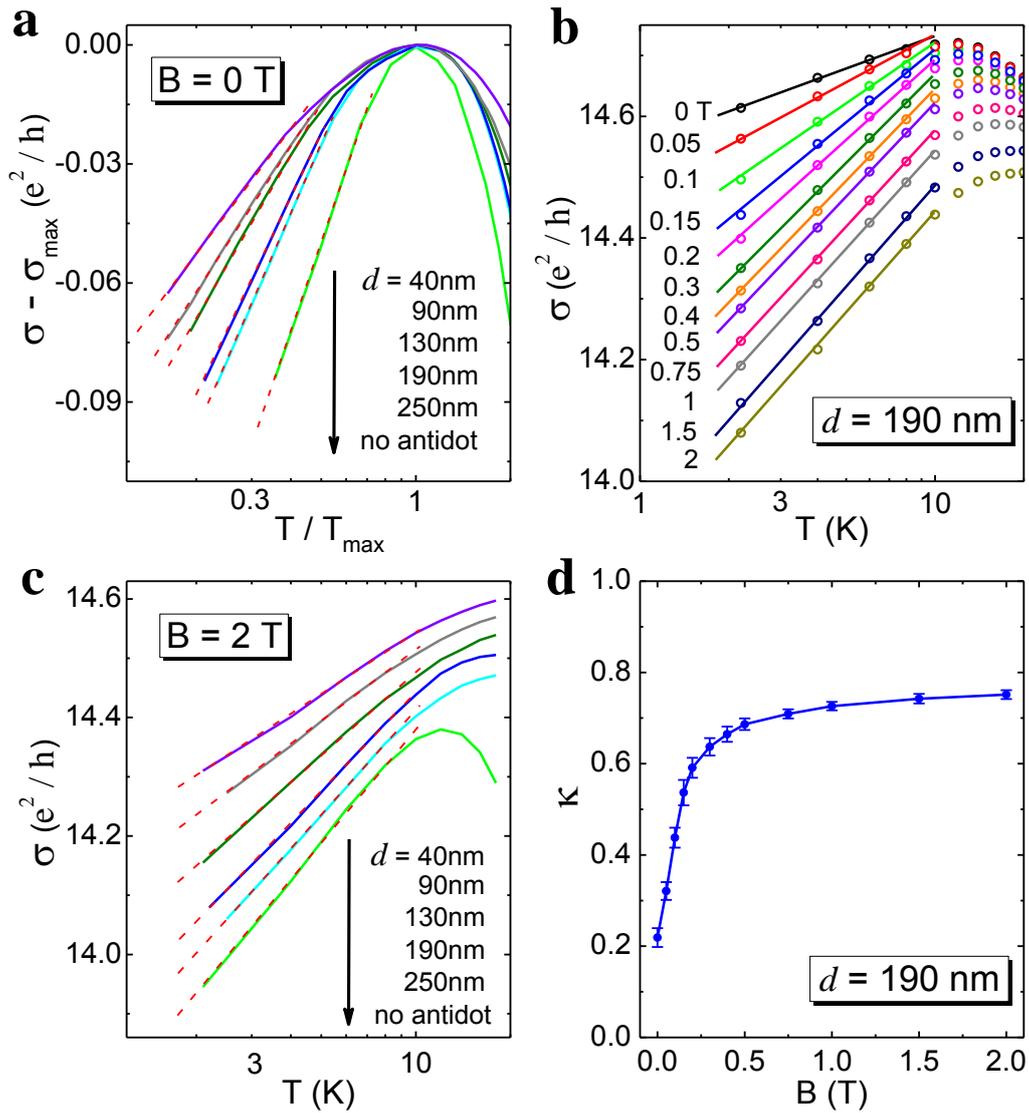

Figure 2. (a) Measured logarithmic temperature-dependent conductivity (solid lines) at zero magnetic field for all samples with the edge-to-edge distance between antidots $d$ indicated. The red dashed lines are guide for eyes. (b) Measured logarithmic temperature-dependent conductivity at indicating perpendicular magnetic fields (open circles) for sample d190. Solid lines are the linear fitting curves. (c) Measured temperature-dependent conductivity (solid lines) at $B = 2$ T for all samples with $d$ indicated. The curves are offset for clarity and the red dashed lines are guide for eyes. (d) The slope obtained by linear fitting the curves in (b) is plotted (solid symbols) as a function of magnetic field for sample d190. Solid lines are guide for eyes.

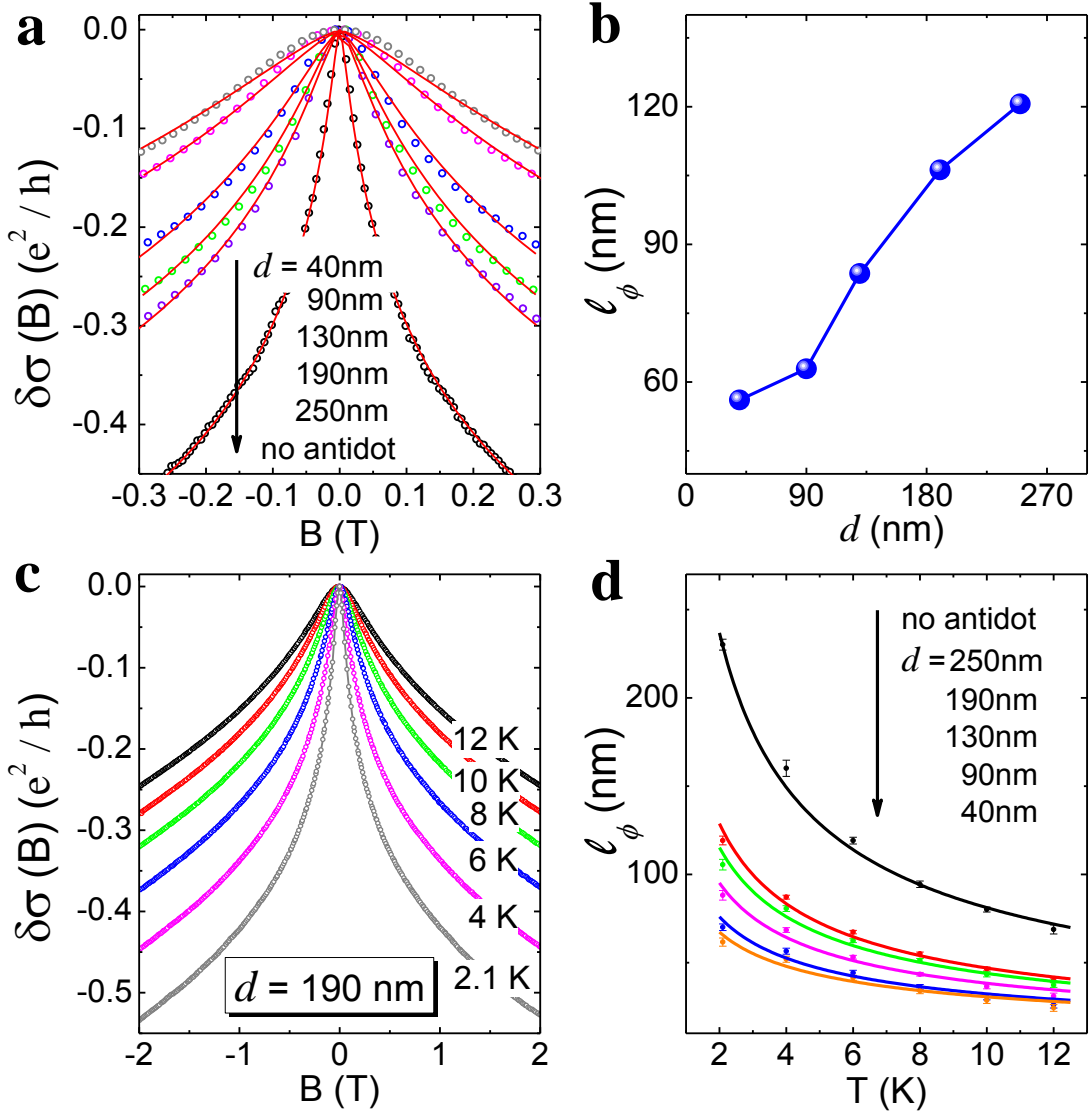

Figure 3. (a) Measured (open symbols) and fitted (solid curves) magnetoconductivity curves of all samples at 2.1 K with the magnetic field $B$ perpendicular to the sample surface. (b) $\ell_\phi$ as a function of $d$ obtained by fitting the magnetoconductivity curves in (a).[27,38] (c) Measured magnetoconductivity curves of sample d190 at different temperatures. (d) The phase coherence length $\ell_\phi$ vs. temperature $T$. Solid symbols are obtained by fitting magneto-conductivity curves at different temperatures for all samples. Solid curves are the fittings to the solid symbols with $\ell_\phi \propto T^{-p/2}$, and the fitted exponent $p/2$ are given in Table 1.

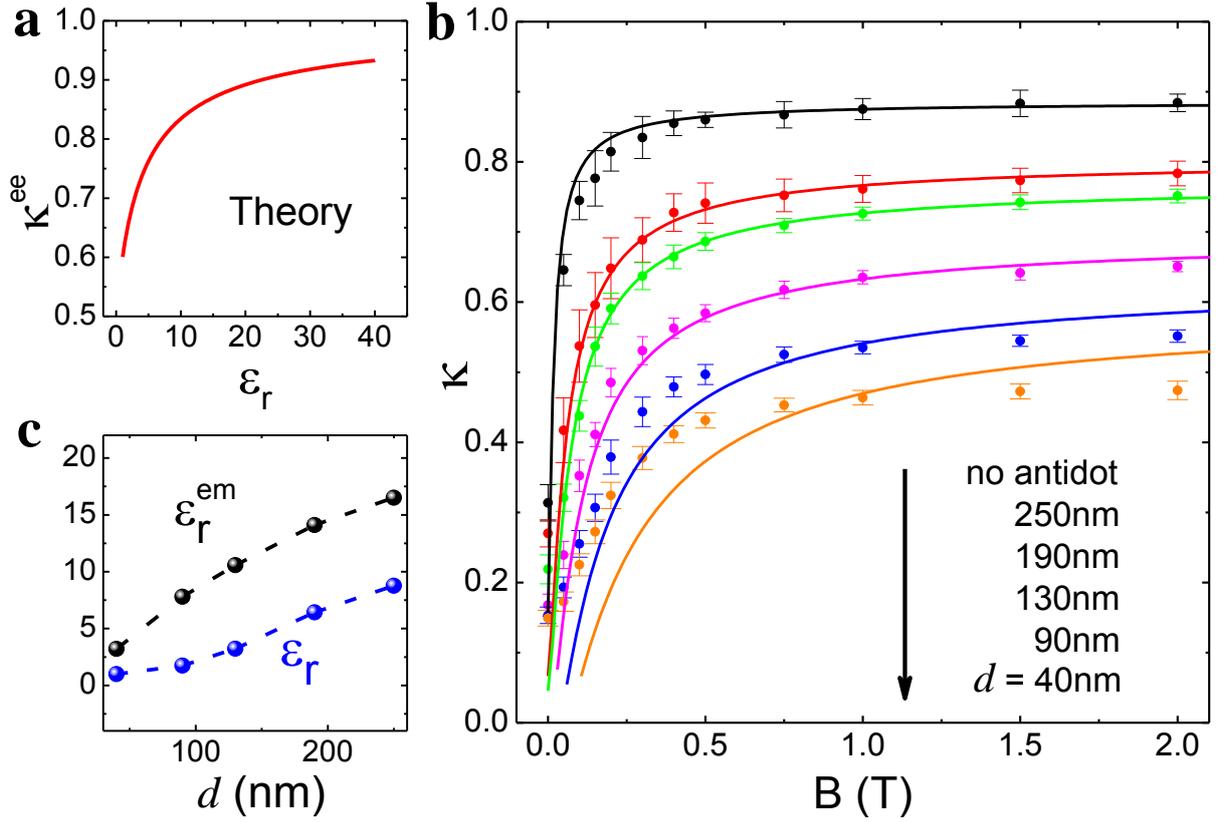

Figure 4. (a) The calculated slope $\kappa^{ee}$ as a function of relative dielectric constant $\varepsilon_r$ for gapless Dirac fermions. $v\hbar$ is chosen to be 2.5 eV · Å, comparable with those in the experiment.[16] (b) The slope $\kappa$ of conductivity-ln$T$ curve as a function of magnetic field $B$ for different antidot distance $d$. The measured $\kappa$ (scatters) are fitted by the theory (solid curves).[27,38] (c) The dielectric constant as a function of $d$, obtained by fitting the transport theory ($\varepsilon_r$) and calculated from the Bruggeman effective medium theory ($\varepsilon_r^{em}$)[46], respectively. The intrinsic dielectric constant $\varepsilon_b = 22.0$ of the no-antidot sample is extracted by fitting the transport theory.

# Supplementary Material for "Tunable interaction-induced localization of surface electrons in antidot nanostructured Bi$_2$Te$_3$ thin films"

Hong-Chao Liu[†,#], Hai-Zhou Lu[‡,#], Hong-Tao He[†,§], Baikui Li[†], Shi-Guang Liu[†],

Qing Lin He[†], Gan Wang[†,§], Iam Keong Sou[†], Shun-Qing Shen[‡,*] and Jiannong Wang[†,*]

[†]*Department of Physics, The Hong Kong University of Science and Technology,*

*Clear Water Bay, Hong Kong, China*

[‡]*Department of Physics, The University of Hong Kong,*

*Pokfulam Road, Hong Kong, China*

[§]*Department of Physics, South University of Science and Technology of China,*

*Shenzhen, Guangdong 518055, China*

[#] *These authors contributed equally to this work*

*  *Address correspondence to: sshen@hku.hk , phjwang@ust.hk*

## S1. DATA FITTING

The ln$T$-dependent conductivity consists of two parts, that is $\sigma(B,T) = \sigma^{qi}(B,T) + \sigma^{ee}(B,T)$, according to the conductivity formula derived for interacting and disordered Dirac fermions in topological insulator.[1] The first part is from the quantum interference $\sigma^{qi}(B,T)$, the second part is from the electron-electron interaction $\sigma^{ee}(B,T)$, whose expressions are given as Equation (2) and (3) in Reference 1, and depend on several sample-dependent parameters: mean free path, phase coherence length, the screening factor $F$, the velocity of Dirac fermion $\gamma$, and $\Delta/2E_F$.[1] Despite the coexistence of multiple bands on the Fermi surface, we generalize $\Delta/2E_F$ as an effective fitting parameter to account for the overall contribution, much like how the fitting parameter $\alpha$ in the HLN formula for conventional electrons[2,3] is generalized in various situations. For all samples, the obtained $\Delta/2E_F \rightarrow 0$, corresponding to $\alpha \sim -0.5$ in the HLN formula.

First, the Fermi wave vector $k_F$ is estimated from the sheet carrier density $n$ from the Hall measurement. Then $k_F$ is put into the semi-classical conductivity

$$\sigma^{sc} = \frac{e^2}{h} k_F \ell,$$  (S1)

to estimate the mean free path $\ell$, where $\sigma^{sc}$ is roughly estimated by the temperature and magnetic field insensitive part of the measured conductivity. We use $\gamma = v\hbar = 2.5$ eV·Å estimated from angle-resolved photoemission spectroscopy of $Bi_2Te_3$,[4] where $v$ is the Fermi velocity. With the values of $\ell$, $\gamma$ and a trial value of $F = 0.1$, the phase coherence length $\ell_\phi$ and exponent $p$ in the relation $\ell_\phi \propto T^{-p/2}$ and the value of $\Delta/2E_F$ are fitted from the magnetoconductivity (see Figure 4 in the main text), at different temperatures. Then, with the value of $\ell$, $\gamma$, $\Delta/2E_F$ and the power-law relation $\ell_\phi \propto T^{-p/2}$, the screening factor $F$ is fitted from the slope $\kappa$ of the conductivity vs. $\ln T$ at high magnetic fields (0.5 T < B < 2 T), where the slope is dominantly determined by the interaction. Further, the fitted $F$ is put into the magnetoconductivity for a self-consistence fitting for all the parameters. Finally, the relative dielectric constant $\varepsilon_r$ is calculated from the convergent value of $F$.

## S2. MEAN FREE PATH

For the mean free path, as shown in Table 1, it is shortened with increasing antidot density. The shortening leads to a suppression of the semiclassical conductivity, which is proportional to $\ell$, and shifts of the quantum interference and electron-electron interaction parts of the conductivity, which is proportional to $\ln \ell$. However, these changes do not affect the slope of the conductivity-$\ln T$ curves. In other words, although both the quantum interference and Altshuler-Aronov effect are attributed to the disorder, the slope is a universal quantity that is independent of the strength of disorder scattering directly.

## S3. THRESHOLD TEMPERATURE

In Figure 2a, we also note that the $\ln T$ behavior starts below a threshold temperature $T_{max}$, which is observed to increase with the antidot density. One of the reasons for this change of $T_{max}$ can be understood as follows. The electron-electron interaction dominant $\ln T$ behavior requires that $\ell_T \gg \ell$, where $\ell$ is the mean free path and $\ell_T$ is the thermal diffusion length that is proportional to $1/\sqrt{T}$. With decreasing $d$ of the antidots $\ell$ is reduced, so the regime satisfying $\ell_T \gg \ell$ is also broadened, leading to the increase of $T_{max}$.

## S4. AVERAGE DISTANCE BETWEEN ELECTRONS NEAR THE FERMI SURFACE

The $\ln T$ dependence in the conductivity is given by the electrons near the Fermi surface, as a result of the interplay of the inelastic scattering induced by the electron-electron interaction and elastic scattering by disorder.[1,2] The elastic scattering does not change energy, as a consequence, the interaction is limited to among those electrons near the Fermi surface.

We can estimate average distance between the electrons near the Fermi surface. For massless Dirac fermions, the density of states at the Fermi energy is given by

$$N_F = \frac{E_F}{2\pi(\hbar v)^2} = \frac{k_F}{2\pi\hbar v},$$  (S2)

where $\hbar$ is the reduced Planck's constant, $v$ is the effective velocity, $E_F$ is the Fermi energy measured from the Dirac point, and $k_F$ is the Fermi wave vector. For massless Dirac fermions, $E_F = \hbar v k_F$. The electrons near the Fermi surface are counted by those in an energy interval of $k_B T$, where $k_B$ is the Boltzmann constant and $T$ is the temperature. The density of the electrons near the Fermi surface at the temperature of $T$ is approximately given by

$$n \approx k_B T \frac{k_F}{2\pi\hbar v},$$  (S3)

then the average distance between them is

$$\bar{d} = \frac{1}{\sqrt{n}} \approx \sqrt{\frac{2\pi\hbar v}{k_F k_B T}},$$  (S4)

where $k_B = 8.62 \times 10^{-5}$ eV · K, $\hbar v \approx 2.5$ eV · Å for $Bi_2Te_3$,[4] and $k_F$ is about 0.1 Å$^{-1}$ in our samples according to the Hall measurements. Put the parameters into the above equation,

$$\bar{d} = \sqrt{\frac{2\pi \times 2.5}{0.1 \times 8.62 \times 10^{-5} T}} \approx \frac{135}{\sqrt{T}},$$  (S4)

which is about 40 nm for $T = 10$ K, and 135 nm for $T = 1$ K.